\documentclass[sigconf]{aamas}

\usepackage{paper-2024-aamas-concurrency}

\setcopyright{ifaamas}
\acmConference[AAMAS'24]{Proc.\@ of the 23rd International Conference
on Autonomous Agents and Multiagent Systems (AAMAS 2024)}{May 6 -- 10, 2024}
{Auckland, New Zealand}{N.~Alechina, V.~Dignum, M.~Dastani, J.S.~Sichman (eds.)}
\copyrightyear{2024}
\acmYear{2024}
\acmDOI{}
\acmPrice{}
\acmISBN{}

\acmSubmissionID{1051}

\acrodef{1A1P}{One-Agent-One-Process}
\acrodef{1A1T}{One-Agent-One-Thread}
\acrodef{AA1E}{All-Agents-One-Executor}
\acrodef{AA1EL}{All-Agents-One-Event-Loop}
\acrodef{AA1T}{All-Agents-One-Thread}
\acrodef{AOP}{Agent-Oriented Programming}
\acrodef{API}{Application Programming Interface}
\acrodef{BDI}{Belief-Desire-Intention}
\acrodef{MAS}{Multi-Agent System}
\acrodef{OS}{Operating System}

\title[Concurrency Model of \acs{BDI} Programming Frameworks]{Concurrency Model of \acs{BDI} Programming Frameworks:\\Why Should We Control It?}
\subtitle{Extended Abstract}

\author{Martina Baiardi}
\affiliation{
  \orcid{0009-0001-0799-9166}
  \institution{University of Bologna}
  \city{Cesena}
  \country{Italy}}
\email{m.baiardi@unibo.it}

\author{Samuele Burattini}
\affiliation{
  \orcid{0009-0009-4853-7783}
  \institution{University of Bologna}
  \city{Cesena}
  \country{Italy}}
\email{samuele.burattini@unibo.it}

\author{Giovanni Ciatto}
\affiliation{
  \orcid{0000-0002-1841-8996}
  \institution{University of Bologna}
  \city{Cesena}
  \country{Italy}}
\email{giovanni.ciatto@unibo.it}

\author{Danilo Pianini}
\affiliation{
  \orcid{0000-0002-8392-5409}
  \institution{University of Bologna}
  \city{Cesena}
  \country{Italy}}
\email{danilo.pianini@unibo.it}

\author{Andrea Omicini}
\affiliation{
    \orcid{0000-0002-6655-3869}
    \institution{University of Bologna}
    \city{Cesena}
    \country{Italy}}
\email{andrea.omicini@unibo.it}

\author{Alessandro Ricci}
\affiliation{
  \orcid{0000-0002-9222-5092}
  \institution{University of Bologna}
  \city{Cesena}
  \country{Italy}}
\email{a.ricci@unibo.it}

\begin{abstract}
We provide a taxonomy of concurrency models for \acs{BDI} frameworks,
elicited by analysing state-of-the-art technologies,
and aimed at helping both \acs{BDI} designers and developers in making informed decisions.
Comparison among \acs{BDI} technologies w.r.t.\ concurrency models reveals
heterogeneous support, and low customisability.
\end{abstract}


\makeatletter
\gdef\@copyrightpermission{
    \begin{minipage}{0.3\columnwidth}
        \href{https://creativecommons.org/licenses/by/4.0/}{\includegraphics[width=0.90\textwidth]{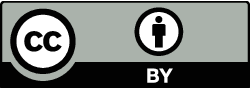}}
    \end{minipage}\hfill
    \begin{minipage}{0.7\columnwidth}
        \href{https://creativecommons.org/licenses/by/4.0/}{This work is licensed under a Creative Commons Attribution International 4.0 License.}
    \end{minipage}
    \vspace{5pt}
}
\makeatother

\begin{document}


\pagestyle{fancy}
\fancyhead{}


\maketitle


\section{Introduction}

By construction,
\ac{BDI} agents are able to carry on multiple intentions at any given time~\cite{Rao1996},
and many research and software-development efforts have been devoted to the definition of \ac{BDI} architectures
and programming languages giving precise semantics to the \emph{concurrent} execution of such intentions~\cite{BordiniHW2007}.

%
As computational entities, agents are autonomous if they encapsulate their own \emph{control flow}~\cite{objag-jot1}.
Control-flow encapsulation is commonly referred to as \emph{computational} autonomy~\cite{artifacts-jaamas17},
and it is considered a necessary
pre-requisite for autonomy in software agents.
On mainstream programming platforms,
autonomy is achieved by mapping each agent onto ad-hoc control-flow-related primitives,
such as threads, processes, or event loops;
providing different trade-offs in terms of efficiency, determinism, and reproducibility of the \acp{MAS} built on top of them.
%
%

Adopting the right concurrency model is essential,
as it deeply impacts many aspects of the agent programming framework
and 
the dynamics of all \acp{MAS} leveraging it.
In particular,
the concurrency model affects whether,
and to what extent, multiple agents can run at the same time,
impacting performance and efficiency of \acp{MAS}
; 
opposedly,
parallelism
as well as the determinism of the overall \ac{MAS} dynamics,
is a strict requirement in applications requiring reproducibility,
such as multi-agent based simulation~\cite{BandiniMV09}.


Dealing with concurrency is commonly acknowledged as error-prone and challenging.
%
Thus,
mainstream programming
platforms provide
dedicated libraries and language constructs
shielding developers
from
the intricacies of concurrency. 
%
Similarly,
\ac{AOP} tools and frameworks 
come with their own concurrency model,
often hidden under the hood.
%

Although hiding concurrency details is
helpful to reduce the learning curve,
experienced developers
unaware of the nuances of the framework they are relying upon
may have reduced control over the execution of their \acp{MAS}
and the trade-offs that come with it.
%
This is particularly true for \ac{BDI} agent technologies,
where the semantics of intention scheduling can be realised in many different ways. 
%
Arguably,
\ac{BDI} technologies should rather let \acp{MAS} developers choose or configure the concurrency model of their systems,
in order to tune the execution of the \ac{MAS} to the specific needs of their application and execution environment.

In this study,
we 
provide a taxonomy of the concurrency models available for \acs{BDI} agent technologies,
and classify several widely used frameworks accordingly.
The current literature on \acs{BDI} agents and concurrency
(e.g.,~\cite{ZatelliRH2015,aloo-agere2013,ricci13-akifest,SilvaML20,Silva2020AnOS})
focuses on agents' \emph{internal} concurrency%
---roughly, how control loops interleave \emph{intentions}.
Conversely,
we focus on \emph{external} concurrency, i.e., the way
multiple agents are mapped onto the
underlying (threads, processes, event loops, executors) concurrency abstractions.
Finally,
we elaborate on the importance of customisable \acp{MAS} execution,
recommending framework designers to promote a neat
separation of the \ac{MAS} specification from its actual runtime concurrency model.

\section{Concurrency models for \acs{BDI} systems}
\label{sec:taxonomy}
\label{sec:concurrency_bdi}

Most modern programming languages support concurrency
through one or more of the following abstractions:
\begin{inlinelist}
    \item \emph{threads},
    the basic units of concurrency~\cite{Dijkstra1965},
    i.e., the executors of
    sequential programs;
    \item \emph{processes}, i.e.,
    containers of threads sharing memory;
    \item \emph{event loops}, i.e.,
    individual threads carrying out 
    sequential programs (tasks) 
    enqueued by users;
    and
    \item \emph{executors}, i.e.,
    event loops with a possibly configurable unbound thread count.
\end{inlinelist}

As introduced,
\emph{external} concurrency models
map \acp{MAS} concepts onto these abstractions;
concretely,
they differ in the way the control loop of each agent is mapped onto them.
Different model provide different granularity:
%
%

\textbf{\ac{1A1T} --}
each agent is mapped onto a single thread,
which is responsible for executing its entire control loop.
The control over the of \ac{MAS} execution is abysmal:
essentially, developers are delegating control to the \ac{OS}.
Determinism is compromised as well,
as the \ac{OS} scheduler may interleave the execution of different agents arbitrarily.
The amount of threads in the \acs{MAS} is unbound,
which may lead to relevant overhead
when the number of active agents (threads) is far greater than the amount of hardware cores/processors. 

\textbf{\ac{AA1T} --}
the whole \acs{MAS} is executed on a single thread
that internally schedules all agents' execution in a custom way,
following some (usually cooperative) scheduling policy.
This model enables fully \emph{deterministic} execution of \acp{MAS},
as parallelism is absent.
Hence, it is desirable when reproducibility is a concern,
such as in testing or reproducible simulations,
but it is unsuitable for performance-critical scenarios,
when hardware capable of parallel computation is available.


\textbf{\ac{AA1EL} --}
the whole \acs{MAS} is executed on a single event loop,
which internally schedules all agents' execution with a first-in-first-out queue of tasks,
guaranteeing fairness by design.
\ac{AA1EL} is equivalent
(also in terms of determinism and performance)
to an \acs{AA1T} strategy
with fair
scheduling (e.g., round-robin). 
%
%


\textbf{\ac{AA1E} --}
%
each agent's activity 
is
enqueued as task
on a shared \emph{executor}.
However,
tasks are executed concurrently (possibly, in parallel).
%
%
\ac{AA1E} is conceptually equivalent to \acs{1A1T},
%
yet technologically preferable
as, by controlling the executor's thread count,
provides finer control on the degree of parallelism.
%
Two specialisations of this model are possible, depending on whether the alive thread count changes with time:
fixed thread pools and variable thread pools.

Further models
can be obtained by (possibly hierarchical) combinations of the aforementioned ones,
obtaining diverse flexibility/controllability trade-offs.
For instance, considering that event loops, executors, and threads are hosted into processes, we can think of:
%

\textbf{\ac{1A1P} --}
  each agent is a process using threads, executors, or event loops for internal concurrency.


\section{Analysis of \acs{BDI} technologies} 

We analyse a selection of open-source and actively maintained \acs{BDI} programming technologies 
to inspect their external concurrency model(s). 
%
We focus on \jason~\cite{BordiniHW2007}, \astra~\cite{CollierRL15}, \jakta~\cite{BaiardiBCP23}, \phidias~\cite{DUrsoLS19}, \spadebdi~\cite{PalancaRCJT22}, \jadex~\cite{PokahrBL2005}.
%
%
In our analysis, for each \acs{BDI} technology, we combine two approaches:
%
we first run a benchmark 
to reveal
how many threads are involved in a 
\ac{MAS} execution;
%
then,
we inspect the source code
and 
documentation 
to understand which concurrency abstractions are used, and
to what extent they are customisable. 

\begin{table}
  \centering
  \caption{
    \acs{BDI} technologies and concurrency models. Meaning of symbols: ``$\checkmark$'' -- supported out of the box;
    %
    %
    ``$\ast$'' -- supports customizations;
    ``$\sim$'' -- we were unable to conclusively confirm or rule out support.
  }
  \label{tab:experiment_results}
  \resizebox{\columnwidth}{!}{
    \begin{tabular}{r||c|c|c|c|c|c}
      \toprule
      \textbf{Model} $\rightarrow$ & \textbf{\acs{1A1T}}  & \textbf{\acs{AA1T}} & \textbf{\acs{AA1EL}} & \textbf{\acs{AA1E}}  & \textbf{\acs{AA1E}} & \textbf{\acs{1A1P}} \\
      \textbf{Tech.} $\downarrow$ & & & & \textbf{fixed}  & \textbf{variable} \\
      \midrule
      \textbf{\jason~\cite{BordiniHW2007}} & $\checkmark$ & $\checkmark\ast$ & $\checkmark$ & $\checkmark$ & $\checkmark\ast$ & $\checkmark\ast$ \\
      \textbf{\astra~\cite{CollierRL15}} & $\checkmark\ast$ & $\checkmark\ast$ & $\checkmark$ & $\checkmark$ & $\checkmark$ & $\sim$ \\
      \textbf{\jakta~\cite{BaiardiBCP23}} & $\checkmark$ & $\checkmark$ & $\checkmark$ & $\checkmark$ & $\checkmark$ & $\checkmark\ast$ \\
      \textbf{\phidias~\cite{DUrsoLS19}} & $\checkmark$ & $\times$ & $\times$ & $\times$ & $\times$ & $\checkmark$ \\
      \textbf{\spadebdi~\cite{PalancaRCJT22}} & $\times$ & $\times$ & $\checkmark$ & $\times$ & $\times$ & $\checkmark$ \\
      \textbf{\jadex~\cite{PokahrBL2005}} & $\times$ & $\checkmark$ & $\times$ & $\times$ & $\checkmark$ & $\times$ \\
      \bottomrule
    \end{tabular}
  }
\end{table}
\Cref{tab:experiment_results} summarises the results of our analysis
including the \acs{1A1P} model,
which is the basis for agents not sharing memory,
hence,
potentially distributable.
%

\section{Discussion and Conclusion}\label{sec:discussion}

The
concurrency model is a paramount dimension to consider when designing or using a (\ac{BDI}) \ac{MAS} technology.
Generally,
choice is desirable,
as different applications and execution environments may benefit from different concurrency models.


From an application development perspective,
the concurrency models impact primarily reproducibility and performance.
Reproducibility requires determinism
(especially when testing),
supported by \ac{AA1T};
sheer performance is usually better with parallel models like
\ac{1A1T}
or,
preferably,
\ac{AA1E}. 
Some scenarios may be better tackled through
custom concurrency models,
hence,
we recommend \ac{BDI} technology designers to provide dedicated \acsp{API}.

We argue that flexibility in the choice of concurrency models
is a central feature for \ac{BDI} technologies
Thus,
we recommend considering them early in \ac{BDI} framework design:
adopting a specific concurrency model early on may complicate or prevent changing it later.
When support for multiple (customisable) concurrency models is not feasible,
early analysis can still prove beneficial.
For instance, despite being conceptually akin,
\ac{AA1E} is preferable over \ac{1A1T},
as the former supports controlling the overall thread count.

Careful design of the \ac{BDI} framework architecture is essential
to ensure separation between
the \ac{MAS} specification and its actual runtime concurrency model:
the former should be written once,
and the latter should be selected as late as possible
(ideally, at application launch).
Flexibility enables:
\begin{inlinelist}
  \item controlling reproducibility for debugging or simulation,
  \item maximising performance in production,
  \item comparing and selecting the best model for the scenario at hand.
\end{inlinelist}



Summarising,
external concurrency of \ac{BDI} agents is paramount in \ac{MAS} engineering,
%
%
Yet, we believe further investigation is needed to
provide a general concurrency blueprint for \ac{BDI} technologies.


\begin{acks}
This work has been partially supported by:
\begin{inlinelist}
    \item ``WOOD4.0'',
        Emilia-Romagna project, art. 6 L.R. N. 14/2014, call 2022;
    \item ``FAIR'',
        (PNRR, M4C2, Investimento 1.3, Spoke 8, P.E. PE00000013);
    \item ``2023 PhD scholarship co-funded by NextGenerationEU and AUSL Romagna''
    (PNRR M4C2, Investimento 3.3, D.M. 352/2022);
    and
    \item ``ENGINES'', (Italian MUR PRIN 2022, grant 20229ZXBZM).
\end{inlinelist}
\end{acks}

\newpage

\bibliographystyle{ACM-Reference-Format}
\bibliography{biblio}

\end{document}